# The Implementation of Quantum Computing Operations with Phase Modulated Microwave Pulses and High-Spin Ions at Zero Magnetic Fields

Arifullin M. R., Berdinskiy V. L.

[1,a,b] Orenburg University, 460018, Orenburg, Pobedy av., 13, Russia

[a)]Corresponding author: arifullinm@mail.ru
[b)]vberdinskiy@yandex.ru

**Abstract.** New physical implementations of quantum computing elementary operations by pulse manipulations with electron spins of paramagnetic ions having two electrons and spin S=1 in a zero magnetic field are proposed. New type of microwave pulses with variable phases and different polarizations are able to induce different transitions between ion spin states, and those transitions are described by quantum computing operators. To describe the effects of such pulses, a new way to present exponential operators as polynomials of Pauli matrices is proposed.

## INTRODUCTION

Recently, quantum computer science and quantum computing have achieved significant, but mainly theoretical success. Quantum algorithms have been developed, and different options for the architecture of quantum computers and the various physical implementations of qubits [1], qutrits and other elements of higher order have been suggested. There are many different proposals for the physical realization of a quantum computer [2] - [11], and they are determined by physics of qubits and nature of interactions between them. As examples polymer systems [12], the representation of a qubit by magnetic flux through a superconducting circuit [13], or the use of electrons as qubits on the surface of liquid helium and some others may be mentioned. However, the practical implementation of many proposed physical systems for processing quantum information require very expensive, complex and, as a rule, unreliable physical and technical methods, such as ultra-low temperatures, super-strong magnetic fields, ultra-deep vacuum, complex laser-optical cooling methods. Many physical systems are characterized by strong interactions with the environments, which lead to fast decoherence of quantum states. The superconducting version of the quantum computer with Josephson transitions [14, 15], despite the already existing achievements in the implementation of a separate qubit, has a number of difficulties. They are connected with the necessity of strict control over the perfection of fabrication of tunnel Josephson junctions, temporal characteristics of pulsed impacts, fluctuations of the voltages, which are the main cause of decoherence. All of these schemes are complex and probably not suitable for a scalable quantum computer. Pulsed laser methods of quantum information processing are based on the absorption and re-emission of optical quanta that transmit high energy to the physical system, which will limit the speed and miniaturization of optical quantum computers. Another schemes of quantum computing are based on use of nuclear magnetic resonance effects, however, but those methods have a number of significant problems and can't be used as a fully scalable quantum computer, as far as a number of coupled qubits can't exceed two dozen due to the exponential decrease in the intensity of the measured signal.

Therefore, solid-state quantum computers are considered to be the most promising. This is an area of active researches and is expected to be easier to scale a solid-state system to a large number of qubits. One option is to use the charge and spin of the electron in quantum dots [16], but the use of the charge as a qubit leads to very small coherence times (10-10-10-13 seconds.) due to the strong interaction with the environment and potential fluctuations created by external charges.

In this paper, the use of paramagnetic ions with the total electron spin S = 1 and the zero nuclear spin in a simple cubic or tetragonal crystal lattice as elements of a quantum computer is proposed. Even in the absence of an external magnetic field (in the zero magnetic field), the dipole-dipole and spin-orbit interactions eliminate the degeneration of levels with the total spin SZ = 0 and SZ = ±1 [17, 18]. The level splitting at zero magnetic fields allows abandoning the use of strong magnetic fields for the realization of spin qubits or qutrits. Depending on the type of ions and the crystal lattice, splitting at the zero fields usually falls within the technically available radio-frequency or microwave ranges of the electromagnetic field. The magnetic component of such fields can induce transitions between energy levels with different SZ values. Previously, crystals with impurity high-spin ions in a zero magnetic field have been proposed as active media for microwave masers [19]. The creation of such masers was accompanied by the development of radio engineering methods for controlling spin states of ions. Such ions as Ni2+, Fe3+, Cr2+, Gd3+ and other lanthanides with different sets of spin and energy levels can work as active ions in paramagnetic masers. It is obvious that the use of low temperatures allow to control the initial populations of spin sublevels and spin relaxation and decoherence times.

The aim of this work is to describe the pulse sequences of new type of microwave and radio frequency pulse fields capable to produce some elementary operations of quantum computing algorithms using high-spin paramagnetic ions at a zero magnetic field.

## SPIN DYNAMICS OF HIGH-SPIN IONS

Spin state of paramagnetic ions in tetragonal crystal lattice with anisotropy of "light axis" type at zero magnetic field is determined by spin Hamiltonian

$$H = \tfrac{1}{2} D \cdot S_z^2 ,\qquad(1)$$

where D is the zero field splitting parameter, which is determined by the spin-orbital interaction. For example, the zero field splitting of the $Cr^{3+}$ ion in the crystal lattice $K_3Co(CN)_6$ corresponds to a frequency of 4.98 GHz; active ion $Fe^{3+}$ has 6 energy levels and the zero field splitting corresponds to the frequency 12 - 20 GHz; the Ni2+ ion in the crystal lattice of sapphire $Al_2O_3$ has the splitting D = 26.24 GHz or D = 42 GHz in cadmium chloride $CdCl_2$ [20].

The spin state of ions with two uncoupled electrons and total electron spin S = 1 is the triplet one. Due to the spin-orbit interaction and the crystal field, the sublevels of the triplet state possess different energies and split even in the absence of the magnetic field and Zeeman interactions. Splitting at the zero field causes the state $|T_0\rangle = 2^{-1/2}|\alpha_i\beta_j + \beta_i\alpha_j\rangle = |0\rangle$ lies below the degenerate states $|A\rangle$ and $|B\rangle$, which are formed as superpositions of $|T_+\rangle = |\alpha_1\alpha_2\rangle = |+1\rangle$ and $|T_-\rangle = |\beta_1\beta_2\rangle = |-1\rangle$ states

$$|A\rangle = 2^{-1/2}|\alpha_i\alpha_j - \beta_i\beta_j\rangle = 2^{-1/2}(|1\rangle - |-1\rangle)$$

and

$$|B\rangle = 2^{-1/2}|\alpha_i\alpha_j + \beta_i\beta_j\rangle = 2^{-1/2}(|1\rangle + |-1\rangle).$$

These states are evident to be the well-known Bell's entangled states formed naturally in high-spin paramagnetic ions. It should be noted here that at the zero magnetic field all three spin states $|0\rangle$, $|A\rangle$ and $|B\rangle$ do not have magnetic moments because the average values of all spin operators ($\sigma_{1i} + \sigma_{2i}$) (i = x, y, z) are equal to zero. Therefore, there are no magnetic dipole interactions and additional decoherence mechanism for such ions. These splitted states provide the ability to control logical qubits based on spin states $|0\rangle$, $|A\rangle$ and $|B\rangle$ by using microwave pulses polarized along the OX or OY axes and phase effects similar to geometric Berry's phase effects [21].

The eigenstates of the Hamiltonian (1) are spin states $|0\rangle$, whose energy is E = 0, and two degenerate states $|A\rangle$ and $|B\rangle$, whose energies are E = D. To describe the effects of microwave pulses it is convenient to present the Hamiltonian (1) in the Ising form as the products of Pauli operators $\sigma_{1Z}$ and $\sigma_{2Z}$ for two electron spins S=1/2 and further use the spin Hamiltonian

$$H = \tfrac{1}{2}D \cdot \sigma_{1Z}\sigma_{2Z}. \tag{2}$$

This spin Hamiltonian has the same eigenstates of $|0\rangle$, $|A\rangle$, and $|B\rangle$. However, there is an important difference between spin Hamiltonians (1) and (2); the spin Hamiltonian (2) has an additional singlet spin eigenstate $|S\rangle = 2^{-1/2}|\alpha_1\beta_2 - \beta_1\alpha_2\rangle$. Fortunately, these symmetric and antisymmetric eigenstates implement various irreducible representations of the SU(2) rotation group, thus any rotation operators, describing the actions of microwave pulses on triplet states, are not able to transfer the spin system from symmetric into the antisymmetric singlet state $|S\rangle$.

Representation of the spin Hamiltonian of paramagnetic ions as the Ising one allows to obtain simple and clear expressions for the operators of the rotations around the OX and OY axes which describe transitions between states $|0\rangle$, $|A\rangle$ and $|B\rangle$. If the microwave field is created by Helmholtz coils or a microresonator where the magnetic component $H_1(t)$ is polarized along the OX axis, then the complete Hamiltonian is

$$H = H_0 + H_X(t) = \tfrac{1}{2}D \cdot \sigma_{1Z}\sigma_{2Z} + 2g\beta H_1(\sigma_{1X} + \sigma_{2X})\cos(\omega_G t). \tag{3}$$

If the microwave field $H_1(t)$ is polarized along OY axis and has phase shift δ, then the spin Hamiltonian should be

$$H = H_0 + H_Y(t) = \tfrac{1}{2}D \cdot \sigma_{1Z}\sigma_{2Z} + 2g\beta H_1(\sigma_{1Y} + \sigma_{2Y})\cos(\omega_G t + \delta). \tag{4}$$

In both formula (3) and (4) $2H_1$ is an amplitude of the magnetic component of the microwave electromagnetic field, $\omega_G$ is a frequency of microwave field, $g$ is an electron g-factor, and $\beta$ is the Bohr magneton.

The effect of microwave pulses on paramagnetic ions implies usage of exponential operators $U = \exp(-iHt)$. The Cayley-Hamilton theorem and Sylvester theorem suggest the use of matrix polynomials of high degrees for such operators. This fact complicates practical computations. However, for spin S=1/2 a simple decomposition of exponential matrices in the form of a trigonometric sum of Pauli operators is known. It will be shown below that for spin S = 1 it will be convenient to present exponential matrices of rotation operators as finite trigonometric polynomials whose terms will be sums of products of Pauli matrices.

Using the interaction representation and the property of the operator product $(\sigma_{1Z}\sigma_{2Z})^2 = I_1 I_2$ ($I_i$ - unit operator), for the two-spin system the rotation operator $U_Y(\theta)$ around the OY axis can be presented as:

$$U_Y(\theta,\delta) = \cos^2\frac{\theta}{2} - \sigma_{Y1}\sigma_{Y2} \cdot \sin^2\frac{\theta}{2} - i\sin\frac{\theta}{2}\cos\frac{\theta}{2}(\cdot(\sigma_{Y1}+\sigma_{Y2})\cos\delta + (\sigma_{X1}\cdot\sigma_{Z2}+\sigma_{X2}\cdot\sigma_{Z1})\sin\delta), \tag{5}$$

here $\theta = g\beta H_1 \tau$ is a rotation angle, $\tau$ is a microwave pulse length, and $\omega_G = D$. If the phase shift is equal to zero, $\delta = 0$, then the spin rotation operator $U_Y(\theta)$ around the OY axis takes the form:

$$U_Y(\theta) = \cos^2\frac{\theta}{2} - \sigma_{Y1}\sigma_{Y2} \cdot \sin^2\frac{\theta}{2} - i\sin\frac{\theta}{2}\cos\frac{\theta}{2} \cdot (\sigma_{Y1} + \sigma_{Y2}). \tag{6}$$

Similarly, the spin rotation operator $U_X(\theta)$ around the OX axis takes the form:

$$U_X(\theta,\delta) = \cos^2\frac{\theta}{2} - \sigma_{X1}\sigma_{X2} \cdot \sin^2\frac{\theta}{2} - i\sin\frac{\theta}{2}\cos\frac{\theta}{2}((\sigma_{X1} + \sigma_{X2})\cos\delta + (\sigma_{Y1}\cdot\sigma_{Z2} + \sigma_{Y2}\cdot\sigma_{Z1})\sin\delta). \tag{7}$$

If the phase shift $\delta = 0$, the spin rotation operator $U_X(\theta)$ around the OX axis takes the form:

$$U_X(\theta) = \cos^2\frac{\theta}{2} - \sigma_{X1}\sigma_{X2} \cdot \sin^2\frac{\theta}{2} - i\sin\frac{\theta}{2}\cos\frac{\theta}{2} \cdot (\sigma_{X1} + \sigma_{X2}). \tag{8}$$

All the above mentioned rotation operators are presented as simple sums and productions of one spin Pauli's operators. The operators (5) - (8) take a very simple form for the most common π- and (π/2)-pulses. If the rotation angle $\theta = \pi$, then for pulses with an arbitrary phase shift δ

$$U_Y(\pi,\delta) = -\sigma_{Y1}\sigma_{Y2} \text{ and } U_X(\pi,\delta) = -\sigma_{X1}\sigma_{X2}.$$

If $\theta = \pi/2$, then

$$U_X(\pi/2,\delta) = \frac{1}{2}(1 - \sigma_{X1}\sigma_{X2} - i\cdot((\sigma_{X1} + \sigma_{X2})\cos\delta + (\sigma_{Y1}\cdot\sigma_{Z2} + \sigma_{Y2}\cdot\sigma_{Z1})\sin\delta)).$$

and

$$U_Y(\pi/2,\delta) = \frac{1}{2}(1 - \sigma_{Y1}\sigma_{Y2} - i\cdot((\sigma_{Y1} + \sigma_{Y2})\cos\delta + (\sigma_{X1}\cdot\sigma_{Z2} + \sigma_{X2}\cdot\sigma_{Z1})\sin\delta)).$$

Such representations of rotation operators for spin S = 1 makes its very simple for describing actions of different microwave pulses. This method of constructing the rotation operators is evident can be easily generalized to construct similar operators $U_X(\theta,\delta)$ and $U_Y(\theta,\delta)$ for particles with spin S > 1 [22].

At low temperatures $kT \ll D$, the lower spin state $|0\rangle$ only is populated. Microwave pulses are able to rotate ion electron spins even at zero magnetic fields, and the rotation operators (5) – (8) describe transitions of electron spins from the lower ground state $|0\rangle$ to the state $|A\rangle$, to the state $|B\rangle$, or to a superposition of this states. Pulses polarized along the OX or OY axes act in own two-dimensional subspaces $\{|0\rangle, |A\rangle\}$ or $\{|0\rangle, |B\rangle\}$. For example, microwave pulses polarized along OX axis produce a following superposition of states

$$U_X(\theta,\delta)|0\rangle = \cos\theta|0\rangle - i\sin\theta \cdot e^{-i\delta}|B\rangle. \tag{9}$$

A similar result is obtained for pulses polarized along the OY axis

$$U_Y(\theta,\delta)|0\rangle = \cos\theta|0\rangle - \sin\theta \cdot e^{i\delta}|A\rangle. \tag{10}$$

Both formula (9) and (10) demonstrates that such pulses are able to transform the microwave field phase $\delta$ into the phase shift of spin states $|A\rangle$ or $|B\rangle$.

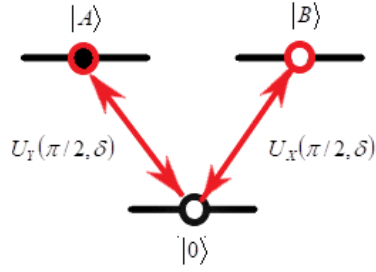

**FIGURE 1.** Transitions between sublevels of paramagnetic ions with spin S = 1 in zero magnetic field under the influence of polarized microwave pulses. The transition $|0\rangle$ - $|B\rangle$ is induced by a π/2 pulse polarized along the OX axis. Transition $|0\rangle$ - $|A\rangle$ is induced by π/2-pulse polarized along the OY axis.

For a microwave pulse polarized along the OX axis, the state $|A\rangle$ is a dark state, and for a microwave pulse polarized along the OY axis, the dark state is $|B\rangle$. Such microwave pulses allow quantum logic operations to be performed, for example, NOT operation. Let the initial state of the paramagnetic ion is the lower ground state $|0\rangle$. The action of the operator $U_X(\theta)$ transfers this state into a superposition

$$U_X(\theta)|0\rangle = \cos\theta|0\rangle - i\sin\theta|B\rangle = c_1|0\rangle + c_2|B\rangle. \qquad (11)$$

Then the pulse $U_X(-\pi/2, \delta)$, acting on the obtained superposition state (11), transfers it to a new state

$$U_X(-\pi/2, \delta)(c_1|0\rangle + c_2|B\rangle) = +c_2(i \cdot e^{-i\delta})|0\rangle + c_1|B\rangle. \qquad (12)$$

If the phase shift of the second pulse is $\delta = \pi/2$, then the pulse $U_X(-\pi/2, \delta)$ simply change coefficients populations c1 and c2 and populations $n_1 = |c_1|^2$ and $n_2 = |c_2|^2$ of $|0\rangle$ and $|B\rangle$ spin states. Thus, the matrix representation of this operator in the basis $|0\rangle, |B\rangle$ has the form

$$U_X(NOT) = \begin{pmatrix} 0 & 1 \\ 1 & 0 \end{pmatrix}, \qquad (13)$$

which is the matrix presentation of the logical NOT operation. Phase shifted microwave pulses are able to produce NOT operations with phase shift.

The phase shift operator S, which is needed for quantum Fourier transformations, can be performed by two pulses polarized, for example, along the OX axis. Such pulses do not affect the state $|A\rangle$ and it is the dark state. If the initial state of the system in the basis $|A\rangle$ and $|0\rangle$ described by the spin vector $|\Psi\rangle = c_1|A\rangle + c_2|0\rangle$.

Sequential application of two (π/2)-pulses $U_X(-\pi/2,0)$ and $U_X(\pi/2,\delta)$ changes the phase factor of the state $|0\rangle$. The first pulse, acting on the ground state $|0\rangle$, transfers electronic spins to the state $|B\rangle$, without changing the state $|A\rangle$,

$$U_X(-\pi/2,0)|\Psi\rangle = c_1|A\rangle + ic_2|B\rangle = |\Psi_1\rangle. \tag{14}$$

The second (π/2, δ) pulse converts the state $|\Psi_1\rangle$ to a new superposition of the initial states $|0\rangle$ and $|A\rangle$, but with a phase shift δ

$$U_X(\pi/2,\delta)|\Psi_1\rangle = c_1|A\rangle + c_2(i\cos(\pi/2)|B\rangle + \sin(\pi/2)\cdot e^{i\delta}|0\rangle) = c_1|A\rangle + c_2 e^{i\delta}|0\rangle. \tag{15}$$

Thus, we obtain the "phase shift" operator S, which is described by the product of $U_X(\pi/2,\delta)$ and $U_X(-\pi/2,0)$ operators. In matrix form in the basis $|A\rangle$ and $|0\rangle$ it has the form

$$S = \begin{pmatrix} 1 & 0 \\ 0 & e^{i\delta} \end{pmatrix}. \tag{16}$$

This logical operation is a key element in the implementation of the quantum Fourier transform.

Rotation of spins around the OZ axis can be useful for the physical implementation of quantum computation algorithms. However, such an operation in a zero magnetic field using pulsed magnetic fields polarized along the OZ axis presents certain technical difficulties and may not be possible. For example, the spin rotation operator $R_Z(\theta)$ S = 1 in two spin representations is as follows

$$R_Z(\theta) = \exp(-i\frac{\theta}{2}(\sigma_{1Z}+\sigma_{2Z})) = \cos^2\frac{\theta}{2} - i\sin\frac{\theta}{2}\cdot\cos\frac{\theta}{2}(\sigma_{1Z}+\sigma_{2Z}) - \sigma_{1Z}\cdot\sigma_{2Z}\sin^2\frac{\theta}{2}. \tag{17}$$

The action of the operator $R_Z(\theta)$ on the state vector $|B\rangle$ transfer it into a superposition of states

$$R_Z(\theta)|B\rangle = \cos\theta|B\rangle - i\sin\theta|A\rangle. \tag{18}$$

The action of the operator $R_Z(\theta)$ on the state vector $|A\rangle$ transforms it into a superposition of states

$$R_Z(\theta)|A\rangle = \cos\theta|A\rangle - i\sin\theta|B\rangle. \tag{19}$$

There is no reason for physical differences of such states for an uniaxial crystal, they do not create a real physical magnetic moment $M_Z$ of the ion. However, such a rotation operation can be carried out by successive application of microwave pulses polarized along the axes OX and OY. For example, the sequential action of operators $U_X(\theta,0)$ and $U_Y(\pi/2,\delta)$ on the state vector $|B\rangle$ translates it into a superposition of states similar to the result of rotation around the axis OZ

$$U_Y(\pi/2,\delta)\cdot U_X(\theta,0)|B\rangle = \cos\theta|B\rangle + i\sin\theta\cdot e^{i\delta}|A\rangle. \tag{20}$$

By changing the polarization of the first and second pulses, a similar transformation of the state vector can be obtained $|A\rangle$

$$U_X(\pi/2,\delta) \cdot U_Y(\theta,0)|A\rangle = \cos\theta|A\rangle - i\sin\theta \cdot e^{-i\delta}|B\rangle. \qquad (21)$$

However, a comparison of formulas (19), (20) and (21) shows the difference between the results of the action of operators $R_Z(\theta)$ (simple rotation around the axis OZ) and operators of microwave pulses. Pulses, characterized by polarization of the magnetic component of the microwave field, duration and phase can additionally change the phase states $|A\rangle$ or $|B\rangle$. The following calculations will show how this additional phase factor affects the physical behavior of spin S = 1.

Let's suppose that initially the spin is in a superposition $|B\rangle$ and $|A\rangle$ of states

$$c_1|B\rangle + c_2|A\rangle = (2^{-1/2})(c_1+c_2)|\alpha\alpha\rangle + (c_2-c_1)|\beta\beta\rangle, \qquad (22)$$

which can be prepared from the initial state $|0\rangle$ by two pulses
$U_Y(\pi/2,\delta) \cdot U_X(\theta,0)|0\rangle = -\cos\theta \cdot e^{i\delta}|A\rangle - i\sin\theta|B\rangle$ where $c_1 = -i\sin\theta$, and $c_2 = -\cos\theta \cdot e^{i\delta}$.

For such superposition of states, the average value of the spin projection $S_z$ and, consequently, the magnetic moment $M_Z$ of the two electron spins of the ion can be different from zero

$$\begin{aligned}\langle S_Z\rangle &= 2^{-1}\langle(c_1+c_2)^*\alpha\alpha + (c_2-c_1)^*\beta\beta|\sigma_{1Z}+\sigma_{2Z}|(c_1+c_2)\alpha\alpha + (c_2-c_1)\beta\beta\rangle = \\ &= 2(c_1c_2^* + c_1^*c_2) = 2\sin(2\theta)\sin\delta\end{aligned} \qquad (23)$$

At any phase shift $\delta \neq 0$ the magnetic moment of the ion $M_Z \neq 0$ and reaches the maximum values at $\delta = \pi/2$ and $\theta = \pi/4$. The effect of the formation of the magnetic moment of paramagnetic ions in the zero magnetic fields by microwave pulses is created by a phase shift of one of the pulses, for example, along the axis OY. The formation of the magnetic moment

$M_Z \neq 0$ from a coherent superposition of spin states is similar to the "alignment-orientation" transition observed in optical experiments [23]. Measurement of the magnetic moment of the $M_Z$ ion makes it possible to identify in which state the spin of electrons is located. Also, the magnetic moment of the $M_Z$ ion can act as a physical measure of the entanglement of the superposition of the Bell states.

Thus, the possibility of realization of logical operations of quantum computing by means of pulse manipulations of high-spin states of paramagnetic ions in a zero magnetic field is shown.

An additional advantage of the scheme of realization of a quantum computer on paramagnetic ions in a zero magnetic field is that in the zero field two degenerate levels ($|B\rangle$ and $|A\rangle$) are the eigenvectors of the spin Hamiltonian and can be protected from decoherence (line broadening), which is essential quantum-logical operations.

## CONCLUSIONS

Phase shifted pulses and paramagnetic ions with total electron spin S = 1 and with zero nuclear spin in a simple cubic or tetragonal crystal lattice in a zero magnetic field are proposed as physical elements of a quantum computer. The advantages of the scheme of quantum calculations with paramagnetic ions in a zero magnetic field have been shown.

For paramagnetic ions with spin S=1, phase shifted microwave pulses capable of converting the spin system into various superposition states of degenerate and non-degenerate levels are proposed.

For spin evolution operators a new presentation of the exponential operators in the form of polynomials Pauli matrices and trigonometric functions of angles of rotation of the spins are derived.

The pulse sequences for NOT operation and the phase shift operator S for quantum Fourier transforms are constructed.

The effect of the formation of the magnetic moment of paramagnetic ions in the zero magnetic field by microwave phase shifted pulses was theoretically predicted.

# ACKNOWLEDGMENTS

The authors gratefully acknowledge the financial support by the Russian Foundation for Basic Research, Grant No 18-37-00374